# Characteristics of Switchable Superconducting Flux Transformer with DC Superconducting Quantum Interference Device

Yoshihiro Shimazu[*] and Takashi Niizeki

*Department of Physics, Yokohama National University, Yokohama 240-8501, Japan*
*CREST, Japan Science and Technology Agency, Kawaguchi, Saitama 332-0012, Japan*

We have investigated the flux transfer characteristics of a switchable flux transformer comprising a superconducting loop and a DC superconducting quantum interference device (DC-SQUID). This system can be used to couple multiple flux qubits with a controllable coupling strength. Its characteristics were measured using a flux input coil and a DC-SQUID for readout coupled to the transformer loop in a dilution refrigerator. The observed characteristics are consistent with the calculation results. We have demonstrated the reversal of the slope of the characteristics and the complete switching off of the transformer, which are useful features for its application as a controllable coupler for flux qubits.
[DOI: 10.1143/JJAP.46.1478]
**KEYWORDS: Josephson junction, superconductivity, DC-SQUID, flux qubit**

## 1. Introduction

Superconducting qubits are a promising candidate for the implementation of a scalable quantum computer.[1] The coupling of qubits is necessary for constructing a multiple-qubit gate. A flux qubit, which is a superconducting loop interrupted by ultrasmall Josephson junctions,[2] can be coupled inductively by using the flux generated by circulating currents in the loop. Previous experiments on coupled flux qubits employed fixed coupling through mutual inductance.[3] It is very desirable for such coupling to be switchable with a fast switching time in order to realize efficient operation for multiple qubits.

Mooij *et al*. have presented a scheme for a switchable flux transformer to meet the above-mentioned requirement.[2] This transformer is a closed superconducting loop that contains two Josephson junctions in parallel, which corresponds to the structure of a DC superconducting quantum interference device (DC-SQUID). Its coupling strength can be varied by changing the magnetic flux in the SQUID loop, which is applied by the current in the control coil adjacent to the SQUID loop. We have investigated this original scheme of the switchable flux transformer both theoretically and experimentally. A switchable flux transformer using a DC-SQUID with a different configuration has recently been studied by Castellano *et al*.[4] The advantage of our scheme for the switchable flux transformer over theirs will be discussed.

Instead of using a control coil as in ref. 4, we injected a current into a segment of the DC-SQUID to change the effective magnetic flux in the SQUID loop. This method is advantageous in that the control current can be small owing to the large kinetic inductance associated with a superconducting wire,[5] thereby minimizing the influence of the control current on the qubits. The observed flux transfer characteristics will be compared with the calculation results.

Our analysis is restricted to the classical regime; however, note that a quantum superposition of two magnetic flux states has been experimentally demonstrated for the same circuit with different design parameters.[6]

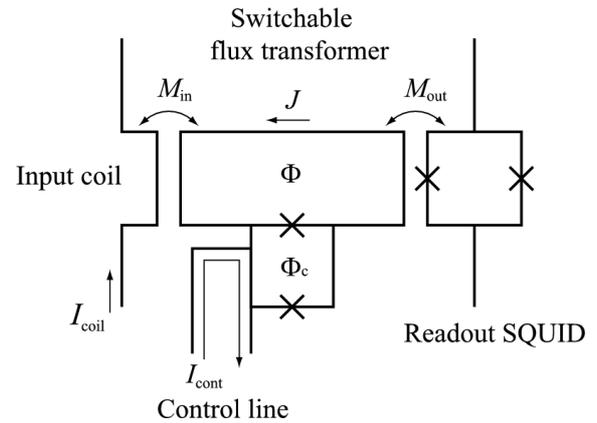

Fig. 1. Schematic of switchable flux transformer. The crosses represent Josephson junctions. An input coil and a DC-SQUID for readout are coupled through the mutual inductances $M_{in}$ and $M_{out}$, respectively, in order to measure the characteristics of the transformer. The transformer involves a controlling DC-SQUID. The flux transfer characteristic is controlled by the magnetic flux $\Phi_c$ in the SQUID loop, which is varied by $I_{cont}$ injected into a loop segment of the SQUID. The magnetic flux produced by the circulating current $J$ is detected by the readout SQUID.

## 2. Theoretical Analysis and Calculation Results

Figure 1 shows the schematic of the switchable flux transformer. An input coil and a DC-SQUID for readout, which are used to measure the characteristics of the transformer, are also shown. We will present the experimental results for the sample described in the schematic. The macroscopic variables that describe the switchable flux transformer are the fluxes $\Phi$ in the main loop (inductance $L$) and $\Phi_c$ in the controlling SQUID loop (inductance $l$).[7] The externally applied fluxes for the loops, $\Phi_x$ and $\Phi_{cx}$, are given by the currents $I_{coil}$ and $I_{cont}$ in the input coil and control line, respectively. We assume that the critical currents $I_0$ of the junctions in the SQUID are equal. The 2D potential describing the system is given by

$$U(\varphi, \varphi_c) = \frac{\Phi_0^2}{4\pi^2 L}\left(\frac{1}{2}(\varphi - \varphi_x)^2 + \frac{1}{2}\gamma(\varphi_c - \varphi_{cx})^2 \right.$$
$$\left. - \beta_0 \cos(\varphi + \frac{\varphi_c}{2})\cos(\frac{\varphi_c}{2})\right), \quad (1)$$

where $\gamma = L/l$, $\beta_0 = 4\pi I_0 L/\Phi_0$, $\varphi = 2\pi\Phi/\Phi_0$, $\varphi_x = 2\pi\Phi_x/\Phi_0$, $\varphi_c = 2\pi\Phi_c/\Phi_0$, $\varphi_{cx} = 2\pi\Phi_{cx}/\Phi_0$, and $\Phi_0$ is the flux quantum.

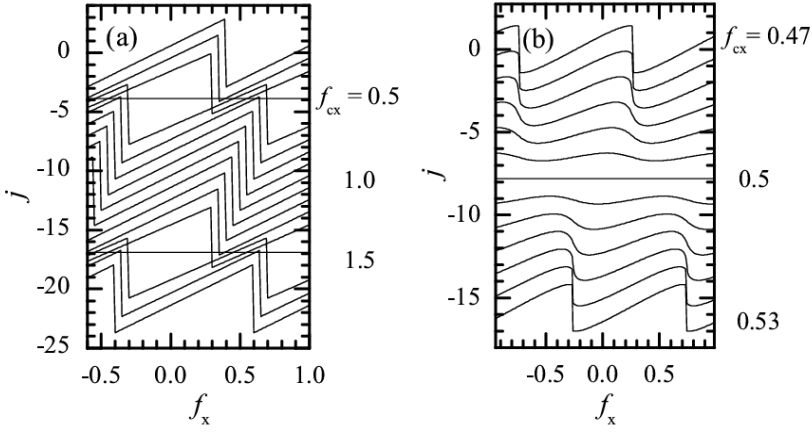

Fig. 2. Calculated circulating current $J$ in the transformer main loop as a function of $f_x$ that is externally applied to the loop. The results for various values of $f_{cx}$ are compared. The curves are offset for better visibility. The range of $f_{cx}$ values is from (a) 0.2 to 1.8 and (b) 0.47 to 0.53.

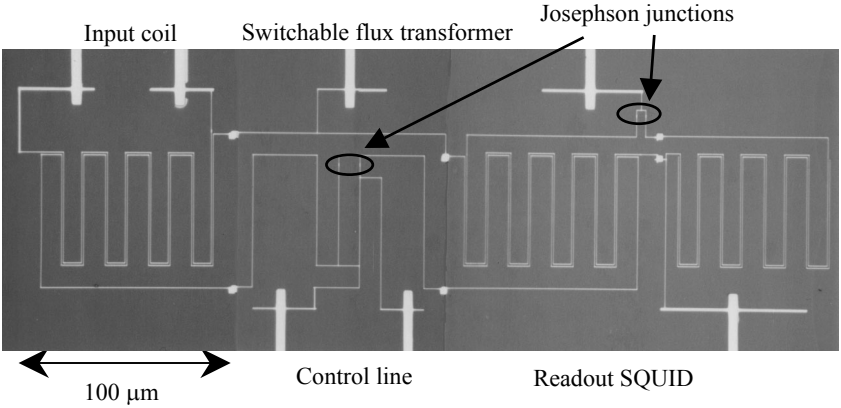

Fig. 3. Optical microscopy image of the sample. All the superconducting wires and Josephson junctions are made of Al. The circles denote the positions of the two Josephson junctions of the DC-SQUIDs.

In the case of $\gamma \gg 1$, similarly to that in our sample, $\Phi_c$ is frozen to the equilibrium value $\Phi_{cx}$. Then, the effective 1D potential for $\varphi$ is given by

$$U(\varphi) = \frac{\Phi_0^2}{4\pi^2 L}\left(\frac{1}{2}(\varphi - \varphi_x)^2 - \beta_0 \cos(\varphi + \frac{\varphi_{cx}}{2})\cos(\frac{\varphi_{cx}}{2})\right). \quad (2)$$

In the classical ground state of the system, $\varphi$ is fixed as the minimum of this 1D potential. The circulating current in the main loop is given by

$$J = \frac{\Phi_0}{2\pi L}(\varphi - \varphi_x). \quad (3)$$

The equation that determines the equilibrium value of $J$ as a function of $\varphi_x$ and $\varphi_{cx}$ is expressed as

$$j + \beta_0 \sin(j + \varphi_x + \frac{\varphi_{cx}}{2})\cos\frac{\varphi_{cx}}{2} = 0, \quad (4)$$

where $j = 2\pi LJ/\Phi_0$. This equation implies that $J$ is a periodic function of $\Phi_x$ and $\Phi_{cx}$ with a period of $\Phi_0$. The variation in $J$ can be detected by the readout DC-SQUID coupled to the main loop of the transformer. The overall flux transfer function is determined by the response of $J$ to $I_{coil}$. This transfer function is also a periodic function of $I_{cont}$, which is proportional to $\Phi_{cx}$, and can be calculated using eq. (4).

Figures 2(a) and 2(b) show the calculated circulating current $j$ in the classical ground state as a function of the normalized input flux $f_x$ for various control fluxes $f_{cx}$, where $f_x = \Phi_x/\Phi_0$ and $f_{cx} = \Phi_{cx}/\Phi_0$. In the calculation, the parameter $\beta_0$ is assumed to be 15, which is relevant to the sample for which experimental data will be presented later. As shown in Fig. 2(a), the phase of the oscillation of $j$ as a function of $f_x$ gradually changes with increasing $f_{cx}$, until $f_{cx}$ becomes a half-integer. It suddenly jumps as $f_{cx}$ crosses the half-integer value; at this point, $j$ is zero and the input flux is not transferred. The characteristics corresponding to $f_{cx}$ being in the close vicinity of a half-integer value is shown in Fig. 2(b). The amplitude of the oscillation is very small, and the phase change is π as $f_{cx}$ crosses the half-integer value.

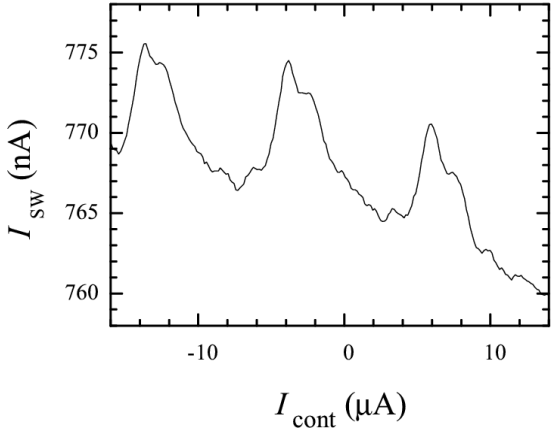

Fig. 4. Switching current $I_{SW}$ of readout DC-SQUID as a function of $I_{cont}$. The current $I_{coil}$ in the input coil was zero, and the external magnetic field was fixed. $I_{SW}$ is modulated periodically with a period of $\Phi_0$ with respect to the flux in the SQUID loop of the transformer.

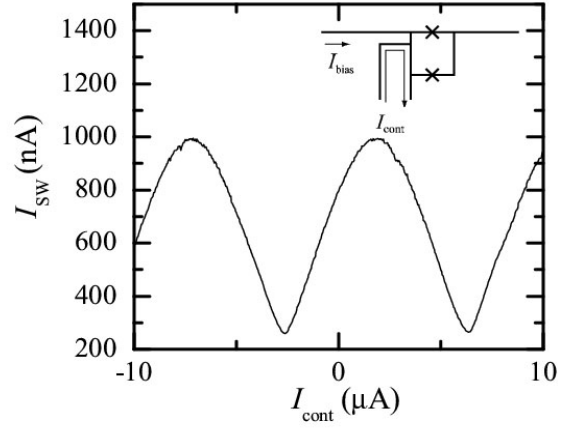

Fig. 6. Dependence of $I_{SW}$ on $I_{cont}$ for sample for which schematic is shown in inset. The switching current $I_{SW}$ was recorded as the bias current $I_{bias}$ was ramped. From the period of the oscillations, the mutual inductance between the control line and the SQUID is determined.

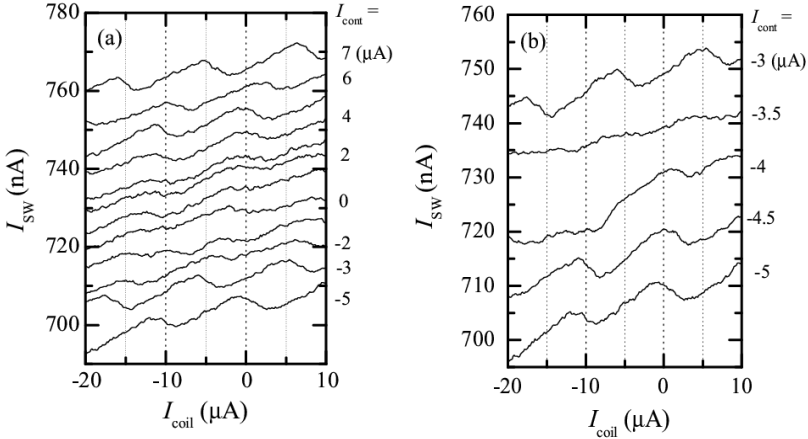

Fig. 5. Switching current $I_{SW}$ of readout SQUID as a function of $I_{coil}$ for various values of $I_{cont}$. The periodicity of the signal corresponds to the flux change in $\Phi_0$ in the main loop of the transformer. The range of $I_{cont}$ values is from (a) –5 to 7 μA and (b) –5 to –3 μA. An abrupt change in the phase of the oscillation is observed for $I_{cont}$ between –4 and –3 μA and that between 6 and 7 μA.

## 3. Experimental Results and Discussion

The sample was fabricated by *e*-beam lithography. Figure 3 shows an optical microscopy image of the sample. All the superconducting wires and Josephson junctions were made of Al. The junctions were fabricated using shadow deposition;[8] their areas were approximately 0.06 (μm)². The critical current $I_0$ of each junction in the DC-SQUID was found to be approximately 500 nA. The area of the readout SQUID loop is 5700 (μm)². The readout SQUID loop was made relatively large in order to obtain a large output flux signal by enhancing its coupling with the transformer. The self inductance $L$ of the transformer main loop is estimated to be 5100 pH. The experiment was carried out in a dilution refrigerator at a base temperature of 25 mK. We measured the switching current $I_{SW}$ of the readout DC-SQUID as a function of the input coil current $I_{coil}$, the control current $I_{cont}$, and an external magnetic field. Note that the change in $I_{SW}$ is proportional to the change in the circulating current in the transformer loop since the change in $I_{SW}$ is considerably smaller than the maximum critical current of the DC-SQUID.

Figure 4 shows $I_{SW}$ as a function of $I_{cont}$ at a fixed magnetic field with $I_{coil} = 0$, while Figs. 5(a) and 5(b) show the dependence of $I_{SW}$ on $I_{coil}$ for various $I_{cont}$ values. The periodic modulations shown in these figures exhibit the periodic behavior of the circulating current in the transformer as a function of both $\Phi_x$ and $\Phi_{cx}$, which was explained in the previous section. The linear background in the figures is due to the spurious coupling of the input coil and control line to the readout SQUID. The opposite tendencies of the backgrounds of Figs. 4 and 5 can be attributed to the directions of $I_{coil}$ and $I_{cont}$ shown in Fig. 1.

From the period of the modulation, the mutual inductance between the input coil and the transformer main loop is estimated to be 180 pH, while that between the transformer SQUID loop and the control current line is determined to be 220 pH. Since the transformer SQUID loop and the control current line share the same line of 40 μm length, the mutual inductance is the sum of the geometrical inductance and kinetic inductance. In order to verify the estimate of the mutual inductance between the transformer SQUID loop and the control current line, we measured this mutual inductance directly in a separate sample.[10] Figure 6 shows the schematic of the sample for this measurement in addition to the observed switching current as a function of $I_{cont}$. The lengths of the lines shared by the control line and the DC-SQUID are the same for these samples. The period of the oscillations of $I_{SW}$ almost agrees with that shown in Fig. 4, which implies that the effective mutual inductances associated with the control line are almost the same for the two samples. The small difference between the mutual inductances of the two samples is attributed to the small difference in the linewidth of the control line.

The observed mutual inductance is in agreement with the sum of the geometrical inductance estimated from the geometry of the sample and the estimated kinetic inductance. The origin of the kinetic inductance is as follows: the phase difference γ along a wire is given by the line integral of the current density $j_s$ as

$$\gamma = \frac{2\pi}{\Phi_0} \int \mu_0 \lambda_L^2 j_s dl, \quad (5)$$

where $\lambda_L$ is the London penetration depth.[9] If one assumes a homogeneous current distribution, the kinetic inductance becomes

$$L_K = \mu_0 \lambda_L^2 \frac{l}{S}, \quad (6)$$

where $l$ is the length and $S$ is the area of the cross section of the wire.[5] In our sample, we have estimated $\lambda_L$ to be approximately 190 nm.[11] The contribution of the kinetic inductance is estimated to be approximately 80% of the total inductance in the present sample. If the control line was separated from the control SQUID, the mutual inductance would be considerably smaller and the required control current would exceed the superconducting critical current of the control line with a width of 200 nm.

In Fig. 5(a), the gradual change and sudden jump in the phase of the oscillation with increasing $I_{cont}$ are indicated, which is consistent with the features shown in Fig. 2(a). The discontinuities in the circulating current, as shown in the calculation, are not observed experimentally. This can be attributed to the thermal excitation and fluctuations in magnetic field. The experimental data in which the normalized magnetic flux in the control SQUID loop $f_{ex}$ is near the half-integer value is shown in Fig. 5(b). The decrease in the amplitude of oscillation and the change in the phase, which qualitatively agree with the calculation shown in Fig. 2(b), are also observed in this figure.

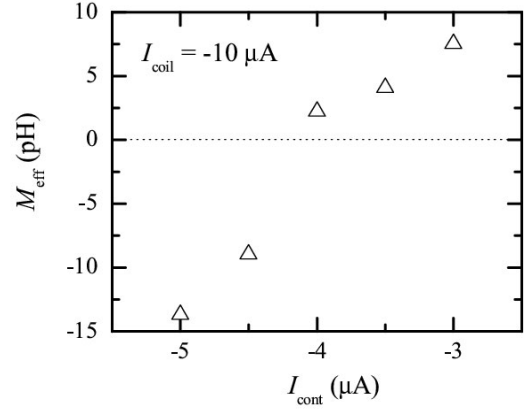

Fig. 7. Effective mutual inductance $M_{eff}$ between input coil and readout SQUID as a function of $I_{cont}$. The overall flux transfer function is expressed by this parameter. The operation point is chosen to be $I_{coil} = -10$ μA.

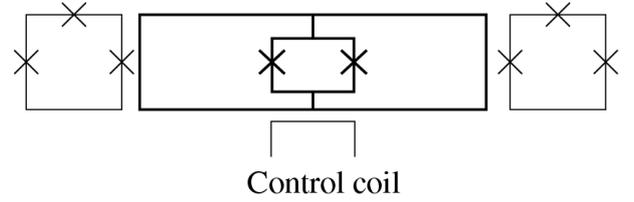

Fig. 8. Schematic of the switchable flux transformer with a different configuration, which is studied in ref. 4.

In the system under investigation, the sign of the slope of the flux transfer characteristics at a particular $I_{coil}$ can be inverted by varying $I_{cont}$, as clearly shown in Fig. 5(b). This is an attractive feature of the present system because a complete switching off of the transformer, that is, a zero response for a small finite change in input flux, can be achieved at a particular operation point. It should be noted that in the application of coupling qubits, the response of the transformer to a small flux change between the different flux states is important. The effective flux transfer function $M_{eff}$ in terms of effective mutual inductance is defined to be

$$M_{eff} = \frac{d\Phi_{SQ}}{dI_{coil}} \quad (7)$$

where $\Phi_{SQ}$ is the input flux for the readout SQUID. Figure 7 shows the dependence of $M_{eff}$ at $I_{coil} = -10$ μA as a function of $I_{cont}$, which was derived from the data shown in Fig. 5(b). Effective mutual inductance varies continuously from −14 to 7 pH through zero. The maximum $M_{eff}$ (the effective mutual inductance in the on-state) is higher than the typical mutual inductance between flux qubits reported in a previous study on coupled qubits.[3] Therefore, this system can be useful in realizing significant coupling between flux qubits.

The flux transfer characteristic of the present system should be compared with that of a similar switchable flux transformer with a DC-SQUID, the schematic of which is shown in Fig. 8.[4] The effective flux transfer function of this system is always positive as shown in Fig. 3 of ref. 4. The present system under investigation is advantageous as a controllable flux transformer over the other scheme in that complete switching off can be realized and the sign of the effective flux transfer function can be inverted. Novel applications of this reversal of the differential characteristics should be examined. In the context of quantum computing, these schemes of a switchable flux transformer using a DC-SQUID as a nondissipative switch are useful for their characteristics of integrability, reliability, and negligible additional decoherence.[4]

This device can be used in coupling not only multiple qubits but also a single qubit to a readout SQUID. The influence of SQUID readout on quantum coherence in a flux qubit should be investigated by means of a controllable coupling between the qubit and the readout SQUID.

## 4. Conclusions

In summary, we have investigated both theoretically and experimentally the flux transfer characteristics of a controllable flux transformer comprising a superconducting loop and a DC-SQUID. The observed characteristics are in qualitative agreement with the calculation results. We have demonstrated that the slope of the flux transfer characteristics can be inverted and that complete switching off can be realized at a particular operation point. This system can be used in realizing controllable coupling between flux qubits and between a flux qubit and a read-out SQUID.